# Collective Excitations, Nambu-Goldstone Modes and Instability of Inhomogeneous Polariton Condensates

Ting-Wei Chen[1], Szu-Cheng Cheng[2,*], Wen-Feng Hsieh[1,3,†]

[1]Institute of Electro-Optical Science and Engineering and Advanced Optoelectronic Technology Center, National Cheng Kung University, Tainan, Taiwan
[2]Department of Physics, Chinese Culture University, Taipei, Taiwan
[3]Department of Photonics and Institute of Electro-Optical Engineering, National Chiao Tung University, Hsinchu, Taiwan

Abstract

We study non-equilibrium microcavity-polariton condensates (MPCs) in a harmonic potential trap theoretically. We calculate and analyze the steady state, collective-excitation modes and instability of MPCs. Within excitation modes, there exist Nambu-Goldstone modes that can reveal the pattern of the spontaneous symmetry breaking of MPCs. Bifurcation of the stable and unstable modes is identified in terms of the pumping power and spot size. The unstable mechanism associated with the inward supercurrent flow is characterized by the existence of a supersonic region within the condensate.

In past years, there have been intensive searches for a new Bose-Einstein condensate in solids. Researchers found such a candidate called the microcavity-polariton condensate (MPC), which has been created from the interaction of cavity photons and confined excitons in the strong-coupling regime [1, 2]. Growing research activities in this MPC can be attributed to the system being intrinsically out-of-equilibrium determined by the dynamical balance between interactions, trapping potentials, pumping and decay [3]. Rich phenomena from non-equilibrium many-body physics are accessible in this system. Many phase-transition signatures from inhomogeneous MPCs, such as spectral and spatial narrowing and first-order coherence, were studied by Balili *et al.* [8]. There still exists superfluidity even though the MPC involves a non-equilibrium dissipative character [4-6]. Due to the continuous pumping and disorders of MPCs, the instability of rotationally symmetric states and vortices appear spontaneously without stirring or rotating MPCs [6, 7]. Non-equilibrium MPCs also show the spontaneously rotational symmetry breaking and are unstable towards the formation of vortices and spontaneous array of vortices without any rotational drive [9-11]. Although the spontaneous symmetry breaking followed by non-equilibrium MPCs has been demonstrated, the properties of spontaneous-symmetry-breaking states still need to be understood and are worthily studied.

Non-equilibrium MPCs have been theoretically investigated by several groups, for instance, Keeling *et al.* studied the slow dynamics of MPCs by eliminating the reservoir dynamics [9, 10], while Wouters and Carusotto couple the dynamics of polaritons from condensate to the reservoir with rate-diffusion equation in homogeneous systems [12]. Both of them concluded the same excitation behavior of diffusive Goldstone modes at low momentum, which is recognized as a unique feature coming from the driven-dissipative systems [13]. In this Letter, we shall apply the complex Gross-Pitaveskii equation (cGPE) to study MPCs with pumping, and decay dynamics in a harmonic potential trap [9, 10]. We study collective excitations and instability of MPCs, and show that the existence of Nambu-Goldstone modes within collective-excitation modes can reveal the pattern of the spontaneous symmetry breaking of the condensate. We also discuss the unstable mechanism by analyzing the competition between the supercurrent flow and sound velocity.

We studied MPCs which were trapped by a harmonic potential with trapping frequency $\omega$ and oscillator length $\lambda = \sqrt{\hbar/m\omega}$, where $\hbar$ and $m$ are Planck's constant and polariton mass, respectively. By choosing the length, time and energy scales in units of $\lambda$, $1/\omega$ and $\hbar\omega$, respectively, and further rescaling the wave function $\psi \rightarrow \sqrt{\hbar\omega/2U}\psi$ by the contact potential $U$, we then obtain the dimensionless cGPE [9]:

$$i\frac{\partial}{\partial t}\psi = [-\frac{1}{2}\nabla^2 + \frac{r^2}{2} + \frac{1}{2}|\psi|^2 + \frac{i}{2}(\alpha - \sigma|\psi|^2)]\psi. \tag{1}$$

Here the net gain $\alpha$ represents the pumping power and $\sigma$, which is fixed ($\sigma = 0.3$) throughout this Letter, is the coefficient of gain saturation.

We assume that the step pumping profile, $\alpha(r) = P[1-\Theta(r-R)]$, is a polar symmetric function of the pumping power $P$ and spot size $R$. Here $\Theta(r)$ is a unit step-function. Before calculating the mode pattern of collective excitations, we have to find the steady state of a MPC by taking $\psi(\mathbf{r},t) = \psi_0(\mathbf{r})e^{-i\mu t}$, where $\mu$ is the chemical potential of the system. The stationary solution $\psi_0(\mathbf{r})$ is given by the Madelung transformation, $\psi_0(\mathbf{r}) = \sqrt{\rho(\mathbf{r})}e^{i\phi(\mathbf{r})}$, where $\rho(\mathbf{r})$ and $\phi(\mathbf{r})$ are the density and phase of a MPC, respectively. If we define the supercurrent $\mathbf{v}$ as the gradient of the phase function $\nabla\phi$, i.e., $\mathbf{v}(\mathbf{r}) = \nabla\phi$, we obtain a continuity equation and the Bernoulli's equation from Eq. (1):

$$\nabla \bullet (\rho\mathbf{v}) = (\alpha(r) - \sigma\rho)\rho, \tag{2}$$

$$\frac{1}{2}(r^2 + \rho) + \frac{1}{2}|\mathbf{v}|^2 - \frac{1}{2\sqrt{\rho}}\nabla^2(\sqrt{\rho}) = \mu. \tag{3}$$

If the pumping power is uniformly distributed, the quantum pressure, $\nabla^2(\sqrt{\rho(r)})/\sqrt{\rho(r)}$, and the supercurrent have a small effect on the density distribution; we obtain the Thomas-Fermi (TF) density distribution $\rho(r) = a^2 - r^2$, where the TF radius $a = \sqrt{3P/2\sigma}$ is determined by the balance of net gain and loss of Eq. (2). The rotationally symmetric supercurrent $v(r) = |\mathbf{v}(\mathbf{r})|$ of the TF approximation is given by $v(r) = -Pr(a^2 - r^2)/4a^2$. To compare the supercurrent with the sound velocity $c$, where $c = \sqrt{\rho/2}$, we define a velocity-comparison factor $f$, $f = 1 - (v/c)^2$. For $P = 4.4$ and $a = 4.69$, the TF density, supercurrent profile, and velocity-comparison factor are shown by red dash-dotted lines in (a), (b) and (c) of Fig. 1, respectively. Note that the TF solution does not describe the system correctly. It just gives us crude physical properties of MPCs.

To find the accurate steady-state of MPCs, we solve equations (2) and (3) numerically by applying the shooting method [14, 15] on the 4$^{th}$ order Runge-Kutta integration with proper boundary conditions. The steady-state densities, supercurrents, and velocity-comparison factors of MPCs are shown in Fig. 1 for $R = 2$ (green dotted lines) and $R = 8$ (blue solid lines) under $P = 4.4$. As $R$ increases, the size and density of the condensate also increase and the peaks of density become sharper (see Fig. 1(a)). The density profiles, which differ from the TF density profile, also contain some suppressions of density in the middle region of the condensate. The reason that the steady-state solutions of MPCs deviate from the TF solution is due to the non-equilibrium characters from pumping and decay dynamics of MPCs.

Such non-equilibrium characters create a nonzero supercurrent in the steady state of the system. So the density distribution of the system is affected by the supercurrent: the higher/lower density at a position corresponding to the smaller/larger supercurrent at this point [9]. It is interesting to notice that the variation of the pumping power is equivalent to the effect of changing the pumping spot size. When we decrease/increase the pumping power, the condensate cloud shrinks/grows and the pumping spot size becomes larger/smaller than the condensate size due to no net gain on the edge and the expansion of the condensate. In addition, due to repulsive interactions of polaritons, the chemical potential of the condensate goes higher and creates a blue shift on the total energy as a high density of polaritons has been injected into the system by raising the pumping power [16].

In Fig. 1(b), we show supercurrents of steady states of MPCs. Velocity flows of non-equilibrium MPCs are not zero and depend on the pumping spot size and power. The occurring mechanism of these non-equilibrium flows can be found easily from the continuity equation of Eq. (2). If the density is low, then the gain dominates; however, if the density is high, then the decay term dominates, and particles in the high-density region flow toward the low-density region. Consequently, the velocity flow must exist in between the high- and low-density regions, and is lower/higher with respect to the higher/lower density at a point. From Eq. (3), we can understand the correlated mechanism between the condensate density and corresponding supercurrent. When the pumping is not very strong, we can neglect the quantum pressure in Eq. (3). Supercurrents are zero outside the condensate and at the center of the trap. There must exist a maximal velocity somewhere in the middle region of the condensate, where the density has to be lowered in order to preserve the chemical potential. So the radial position of the maximal supercurrent coincide with the position of density depletion. Indeed, we find a maximal flow velocity in the steady-state solution of a MPC. The maximal velocity of the supercurrent happens at the density suppression and increases with increasing $R$ as shown in Fig. 1(b).

Above some critical pumping spot size $R_c$, there is a supersonic region, where the flow velocity is larger than the sound velocity with the velocity-comparison factor $f < 0$, in the supercurrent profile of MPCs, e.g., r > 1.5 for R = 8 as blue solid line in Fig. 1(c). If the supersonic phenomenon happens within the condensate, the condensate will generate rotationally symmetric breaking states, rotons and vortices; and the condensate becomes unstable in the Landau instability [17, 18] and Feynman theory of superfluidity [19] breaking down. We also observed the flow direction of the supercurrent depending on the pumping spot size: the outward and inward flows correspond to the smaller and larger pumping spot sizes, respectively. In a trap, a small perturbation on the condensate can change its original flow along the radial direction into the azimuthal direction spontaneously. The system then becomes spontaneously rotational symmetry breaking and will contain many degenerate states labeled by the quantum number $\ell$ of angular momentum. According to the Goldstone's theorem, excitation modes of a spontaneous-symmetry-breaking system will have gapless modes, called the Nambu-Goldstone modes [20]. The Nambu-Goldstone modes are the states with dominant degrees of freedom that the system will behave at low energy or long wave length [20]. Therefore it is fruitful that we find the excitation-mode pattern of MPCs in order to understand the property of spontaneously rotational symmetry breaking occurring in MPCs.

By using a perturbation method [21] to calculate the collective-excitation spectra of MPCs in a trap, we investigate the spontaneous-symmetry-breaking properties of MPCs from the Nambu-Goldstone modes that occur in the collective-excitation spectrum, and the stability of MPCs. To study the dynamical properties of MPCs, we consider a small fluctuation $\delta\psi(\mathbf{r}, t) = e^{-i\mu t}[u(\mathbf{r})e^{-i\Omega t} - w^*(\mathbf{r})e^{i\Omega t}]$ on the steady state, where $\Omega$ is the excitation frequency of the system. Substituting the total wave function $\psi(\mathbf{r}, t) = \psi_0(\mathbf{r}) + \delta\psi(\mathbf{r}, t)$ into Eq. (1) and linearizing it, we obtain a pair of Bogoliubov's equations [22], for excitations. Because of the rotational invariance of the condensate, $u(\mathbf{r})$ and $w(\mathbf{r})$ of Bogoliubov's equations may be chosen as these forms in the polar-coordinate: $u(\mathbf{r}) = U(r)e^{i\phi(r)}e^{\pm i\ell\theta}$ and $w(\mathbf{r}) = W(r)e^{-i\phi(r)}e^{\pm i\ell\theta}$, where $\theta$ is the azimuthal angle and $\ell$ is the quantum numbers of angular momentums. The Bogoliubov equations read

$$L[U] - \frac{1}{2}(\rho - i\sigma\rho)W = (\mu + \Omega)U, \qquad (4)$$

$$L^\dagger[W] - \frac{1}{2}(\rho + i\sigma\rho)U = (\mu - \Omega)W, \qquad (5)$$

where

$$L = -\frac{1}{2} e^{-i\phi} \frac{1}{r}\frac{d}{dr}(r\frac{d}{dr}e^{i\phi}) + [\frac{\ell^2}{2r^2} + \frac{r^2}{2} + \rho + \frac{i}{2}\alpha - i\sigma\rho] \quad (6)$$

and its adjoint operator $L^\dagger$ are linear operators acting on $U(r)$ and $W(r)$, respectively. From solving equations (4) and (5) numerically, we find the collective-excitation states and their excitation energies as a function of $\ell$ of the condensate. Because of the non-equilibrium character of MPCs the Bogoliubov equations are non-Hermitian, and excitation frequencies $\Omega$ are complex values [12, 13], whose real parts, Re($\Omega$), and imaginary parts, Im($\Omega$), show excitation energies and decay or growth rates of the condensate, respectively. The decay, Im($\Omega$) < 0, or growth, Im($\Omega$) > 0, behavior of collective-excitation modes indicates the steady state of the condensate is stable or unstable, respectively.

There are many excitation states and we will be mostly interested in the branch with the lowest excitation frequency. The low-lying excitation mode pattern of the condensate are shown in Fig. 2. This mode pattern is quite different for various pumping powers and spot sizes. As shown in Figs. 2(a)-(c), excitation modes of MPCs have Im($\Omega$) < 0 for all $\ell$, when $R \leqq a$, where $a$ = 3.87 (4.69) for $P$ = 3 (4.4). Perturbed MPCs with $R \leqq a$ are stable and will relax back to its stationary state, i.e., be damped, after the pumping power is turned off. The stable excitation modes basically exhibit a trend that the excitation states with the higher angular momentum have the larger numbers of nodes and higher energies. These modes with the lower angular momentum show different behaviors, which can be gapped, linearly dispersing, or diffusive as the long-wavelength modes for spin modes shown in spinor polariton condensates [10]. For condensates with the lower $P$ and smaller $R$, the excitation energies of modes with the lower angular momentum are finite and gapped as shown in Fig. 2(a). These modes, whose decay rates are small, show underdamped rotation. If $P$ goes higher, the decay rates of these modes start increasing and there are gapless modes, Re[$\Omega$] = 0, occurring in the excitation spectra as $R$ becomes larger. In Fig. 2(b), those modes with the linear dispersion relation have a gapless mode, start showing a bifurcation of small decay rates at $\ell = 0$ and exhibit critically damped rotation. There are gapless excitation modes at $\ell = 0$ and 1 in Fig. 2(c). They are diffusive and will show overdamped rotation due to high decay rates. The gapless mode at $\ell = 1$ is a Nambu-Goldstone mode, which indicates the steady-state wave function containing a node and the rotational symmetry of the condensate breaking down spontaneously. The system also tries to restore the continuous rotational symmetry of the condensate according to the Goldstone theorem. Due to the combined effect of radial and rotational currents the steady state with nodes can spontaneously create spiral vortices observed in the previous simulations [9]. We believe that the occurrence of spontaneous vortices in MPCs is a manifestation of spontaneously rotational symmetry breaking of MPCs.

If $R$ is very large, the excitation modes with the higher angular momentum have the higher density distributions in the outskirt of the condensate, and due to a net gain occurring in this region these modes become unstable dynamically, where $\text{Im}(\Omega) > 0$ (see Fig. 2(d)). The value of $\text{Im}(\Omega)$ changes from negative to positive at $\ell = 3$. The excitation spectra exhibit gapless modes not only at $\ell = 0$ and 1 but also at a higher angular momentum, $\ell = 7$. We find that the condensate contains nodes and its symmetry is spontaneously broken from occurring these gapless excitation modes. The condensate with $\ell = 1$, whose $\text{Im}(\Omega) < 0$, generates a stable vortex with a single-flux quantum. The condensate with $\ell = 7$, whose $\text{Im}(\Omega) > 0$, generates a unstable vortex with a multi-flux quantum, and this vortex will break into stable arrays of vortices with a single-flux quantum due to the repulsive effect of fluxes. The existence of Nambu-Goldstone modes with higher angular momentum in the condensate confirm and explain the mechanism of instability forming spontaneous vortex lattices seen by Keeling and Berloff if $R$ is larger than the TF radius [9].

In order to understand the stable and unstable mechanisms of excitation modes, the density and velocity distributions of excitation modes in Fig. 2(d) are compared. We find that the flow velocities of excitation mode at $\ell = 2$ are very small. Then the perturbation of this mode on the condensate will hardly propagate, and the system is stable. On the contrary, the excitation mode at $\ell = 3$ flows inward with large velocity. The fluctuation of this mode on the condensate is propagating fast and enhanced to create the instability of the steady state. We can conclude that the onset of instability is related to the simultaneous inwards flows of excitation quasi-particles with higher angular momentum. To further discuss the instability of MPCs, we plot the phase diagram of MPCs in Fig. 3. MPCs in a trap are stable/unstable as $P$ goes higher/lower and $R$ is smaller/bigger. Although the phase boundary of the TF approximation (dashed line) also shows the same trend, the TF approximation does not match the correct phase boundary (solid line) because the supercurrent of the condensate will affect the density distribution of the steady state significantly, particularly at high pumping power; and the density gradient in the cGPE is actually not negligible as in the TF approximation. The phase boundary determined by the TF approximation is valid as $P$ is lower and $R$ is smaller. As $P$ and $R$ become larger, it is invalid to use the TF radius as the criterion of forming vortex lattices simultaneously [9].

In conclusions, we applied the cGPE to study inhomogeneous MPCs. We obtained the density and supercurrent distributions of steady states of MPCs. The velocity flow at a position is lower/higher with respect to the higher/lower density. Above some critical pumping spot size $R_c$, there is a supersonic region in the supercurrent profile of MPCs. The flow direction corresponding to larger pumping spot sizes is inward. In a trap, a small perturbation on the inward flow can make the system become spontaneously rotational symmetry breaking and will show the existence of Nambu-Goldstone modes within collective-excitation modes. Collective excitations also exhibit a bifurcation of stable and dynamically unstable modes by varying the pumping spot size and power. We explained the mechanism of dynamical instability due to occurring large inward flows of excitations and a supersonic flow in the condensate. In particular, vortices can be generated from unstable Nambu-Goldstone modes with higher angular momentums. Our observations in this Letter are crucial for studying the dynamics of polariton condensates in the future.

We acknowledge the financial support from the National Science Council (NSC) of the Republic of China under Contract No. NSC99-2112-M-034-002-MY3 and NSC99-2112-M-009-009-MY3. S. C. thanks the support of the National Center for Theoretical Sciences of Taiwan during visiting the center.

# References

*sccheng@faculty.pccu.edu.tw

†wfshieh@mail.nctu.edu.tw

[20] Adriaan M. J. Schakel, *Boulevard of Broken Symmetries: Effective Field Theories of Condensed Matter* (World Scientific Publishing, Singapore, 2008).

[21] C. J. Pethick and H. Smith, *Bose-Einstein Condensation in Dilute Gases*, (Cambridge University Press, New York, 2002).

[22] N. N. Bogoliubov, J. Phys. (USSR) **11**, 23 (1947)12

# Figure Captions

Fig. 1. (Color online) Stationary solutions of the condensate. (a) Density distributions, (b) supercurrent distributions, and (c) velocity-comparison factors for the pumping power $P = 4.4$ and pumping spot sizes $R = 8$ (blue solid lines) and $R = 2$ (green dotted lines), respectively. Red dash-dotted lines are the solutions of the Thomas-Fermi approximation.

Fig. 2. (Color online) Collective-excitation spectra versus quantum numbers $\ell$ of angular momentum. Excitation modes of (a) $R = 2$, $P = 3$, (b) $R = 2$, $P = 4.4$, (c) $R = 4$, $P = 4.4$, and (d) $R = 6$, $P = 4.4$, respectively. Red circles and blue triangles show real (left axis) and imaginary (right axis) parts of excitation spectra. Connected lines are used to visualize the trend of excitation spectra.

Fig. 3. (Color online) Phase diagram of the condensate for various pumping spot sizes and powers. Empty blue circles are results from excitation spectra to show the boundary of instability. The dashed green line is the phase boundary from the Thomas-Fermi approximation and the solid blue line is used to visualize the border between stable and unstable modes.

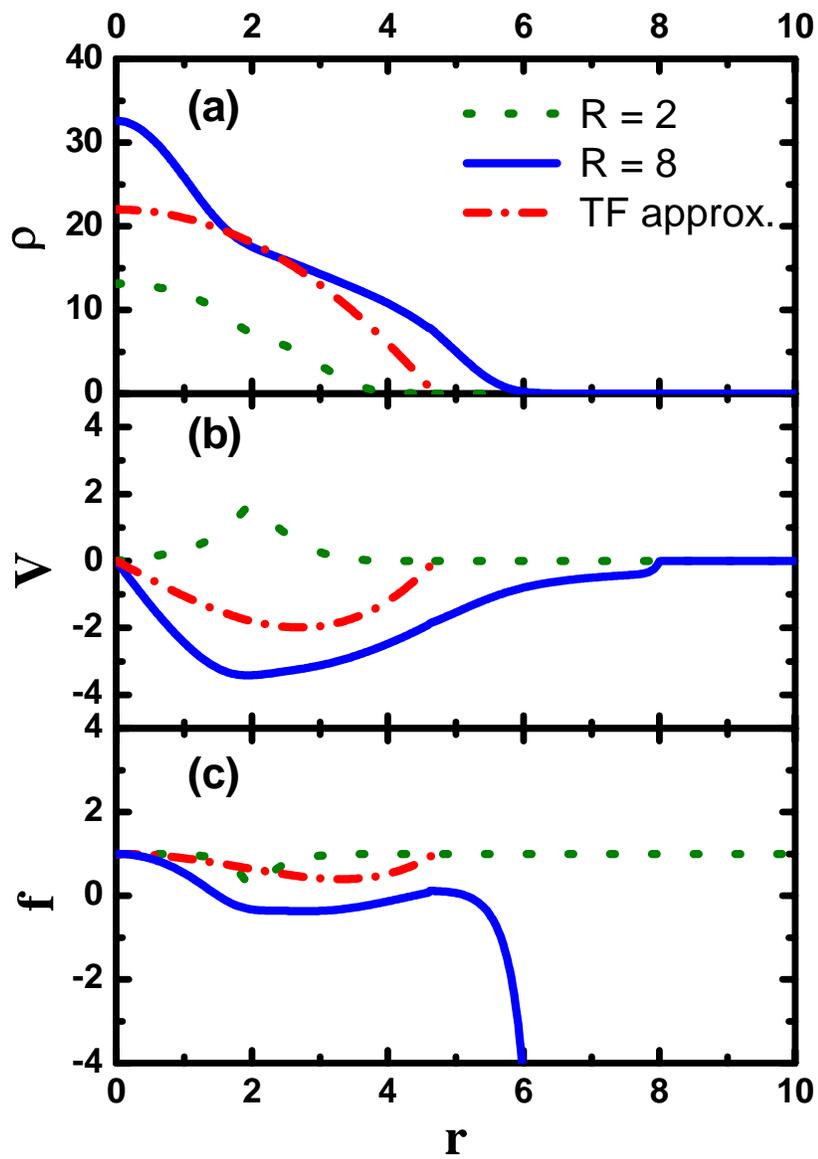

*Fig. 1 (Chen et al.)*

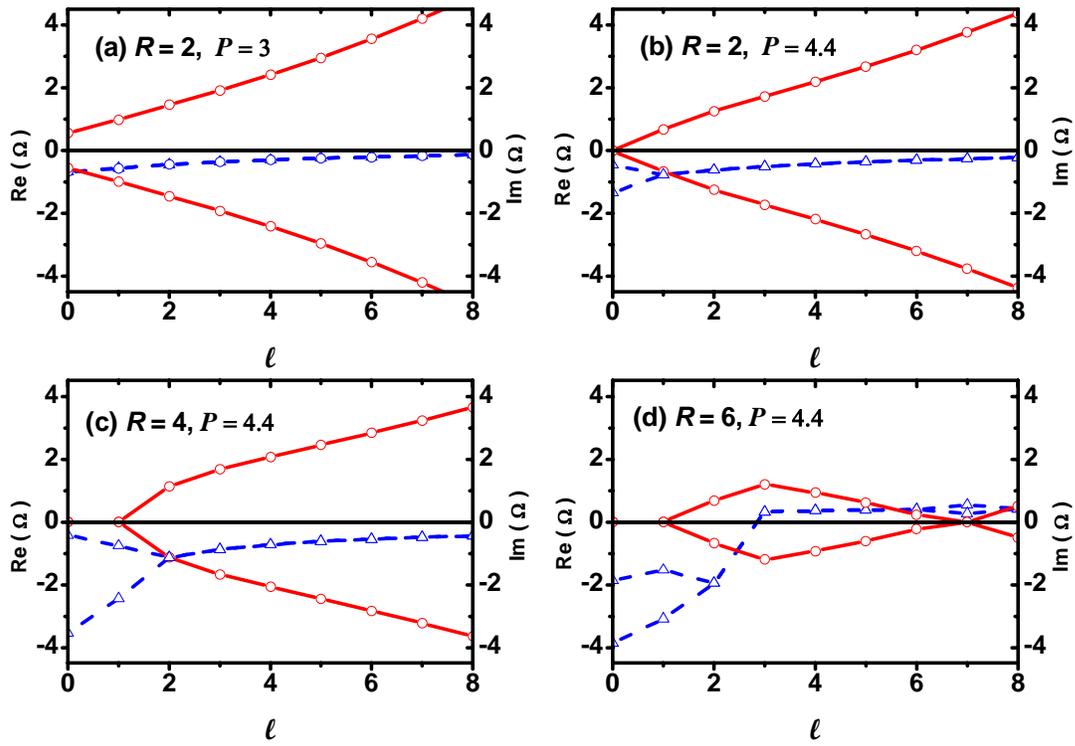

*Fig. 2. (Chen et al.)*

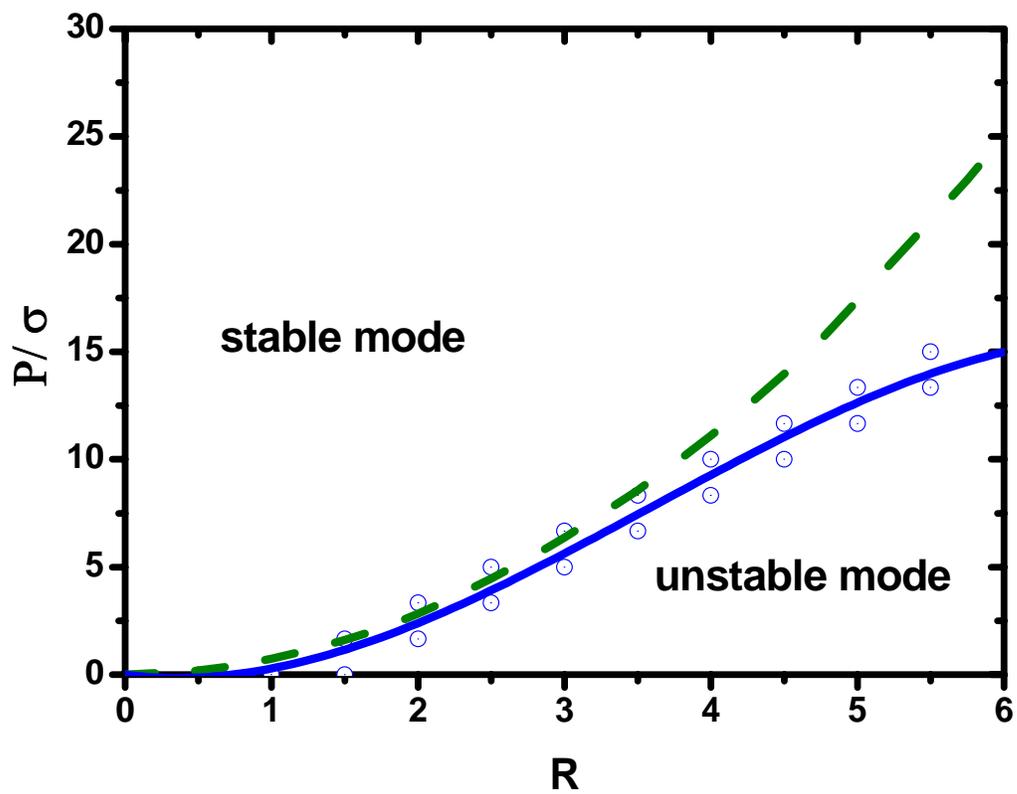

*Fig. 3. (Chen et al.)*